\documentclass{moriond}

\def\be{\begin{equation}}
\def\ee{\end{equation}}
\def\bea{\begin{eqnarray}}
\def\eea{\end{eqnarray}}

\newcommand{\Photo}{\includegraphics[height=35mm]{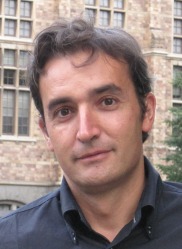}}

\begin{document}
\vspace*{4cm}
\title{Overview of ANTARES results on Dark Matter Searches}

\author{J.D. Zornoza}
\footnote{on behalf of the ANTARES Collaboration}

\address{IFIC (Univ. of Valencia - CSIC, c/Catedr\'{a}tcico Jos\'{e}
  Beltr\'{a}n, 2, 46980, Paterna, Valencia, Spain}

\maketitle\abstracts{
The search for dark matter is one of the most important goals of
neutrino telescopes. The ANTARES detector, installed in the
Mediterranean Sea, is taking data since 2007. Neutrino telescopes have
interesting advantages for this kind of searches, as it will be
discussed. In this talk we will review the status and results of the
analyses performed in ANTARES for several sources: the Sun, Galactic
Centre and the Earth.
}

\section{Introduction}
\label{sec:intro}


There is compelling evidence that a large part of the matter content
of the Universe is made of some something beyond the Standard Model of
particle physics. Among these experimental proofs we can mention the observations from
Planck, the results on the Big Bang
Nucleosynthesis, the rotation curves of
galaxies and the studies of highly red-shifted Ia
supernovae. The conclusion from these experimental
results is that about about 85\% of the matter of the Universe is
non-baryonic.  In the most studied scenarios, it is assumed that
the properties of this matter are the following: weak interaction with
matter, no interaction with photons, non-zero mass and stability. The list of possible candidates for explaining this
component is large, but one of the most appealing frameworks is
SuperSymmetry, in particular in the models which offer neutralino as a
dark matter candidate. The stability of neutralino would be preserved by the
R-parity conservation. 

The search for dark matter is one of the main goals of neutrino
telescopes like ANTARES. The potentially interesting sources include
the Sun, the Galactic Centre, the Earth, dwarf galaxies and galaxy
clusters. Each source has advantages and disavantages. The same
applies to the different detection techniques (direct, indirect,
accelerators), as it will explained.

If dark matter is made of weakly interacting particles (WIMPs), like
neutralinos, they can scatter in astrophysical objects like the Sun or
the Earth and become gravitationally trapped. Their self annihilation
would produce, directly or (more commonly) indirectly high energy
neutrinos. Dark matter forming part of the halo of the Galaxy can also
annhilite and produce a signal, in particular in the direction of the
Galactic Centre.


\section{The ANTARES detector}
\label{sec:antares}

The ANTARES collaboration finished the installation of a neutrino
detector~\cite{antares} in the Mediterranean Sea in 2008. By 2007,
five of the final twelve lines were already taking data. The detector
is deployed at a depth of 2500 m and about 40~km of the French coast,
near Toulon. It consists of 885 photomultipliers installed along
twelve vertical lines anchored to the sea bed. 
The operation principle is based on the detection of the Cherenkov
light induced by the relativistic muons produced after the
charged-current interaction of high energy muon neutrinos close/inside
the detector. 

In the analyses presented in this paper, about 1300 days of data are used,
corresponding to the period from 2007 to 2012. As mentioned before,
only five lines were installed during 2007. 

There are two sources of background. On the one hand, the so-called
atmospheric muons, which are produced by the interaction of cosmic
rays in the atmosphere. This is a huge background that can be
partially reduced by installing the detector at a large depth, as it
is the case for ANTARES. Even at a depth of 2000-2500 m, the
atmospheric muon background is quite important, so only upgoing events
are selected for the analysis, so that the Earth acts as a shield
stopping atmospheric muons. An additional cut in the quality of the events
is also needed in order to reject down-going atmospheric muons which
are badly reconstructed as upgoing. The second kind of background are
the atmospheric neutrinos produced also by cosmic rays. This is an
irreducible background, but expected to be distributed diffusely in
the sky, while the signal would be concentrated around the source.

\section{Analysis method and results}
\label{sec:method}

Different methods have been used for reconstructing the muon
track. For low masses (below $\sim$250~GeV), an algorithm called BBFit has been
used~\cite{bbfit}, which offers a better response for low energies. In
particular, it is used for the events reconstructed with only one
line. For these events, the azimuth information is not available, but
the information on the zenith angle helps to distinguish between
signal and background. For higher masses, a likelihood algorithm is
used~\cite{aafit}.

The analyses presented in this paper have been done using a binned
method. The strategy consists in finding the optimum selection in
terms of neutrino flux sensitivity. The selection variables are the
opening cone angle (i.e. the maximum angular distance to the source)
and a cut in a parameter which describes the quality of the track
reconstruction. All these cuts are chosen following a blind procedure.

\subsection{The Sun}
\label{sec:sun}

For the Sun, the signal is simulated with WIMPSim~\cite{wimpsim}, which takes into
account the main ingredients of the neutrino production and
propagation. Since the compostion of the neutralino is not known, we
consider several channels (pessimistic and optimistic) assuming a
branching ratio of one. The real situation should be between these
extreme cases. One of the advantages of the searches for dark matter
in the Sun, compared with other indirect searches, is that a potential
signal would be free of astrophysical background. Neutrinos from
nuclear reactions in the Sun are of much lower energies. The background
from cosmic rays interacting in the Sun corona is very low. Concerning
the atmospheric background, it can be accurately estimated from scrambled
data. 

\begin{figure}[c]
\begin{center}
\includegraphics[width=0.5\linewidth]{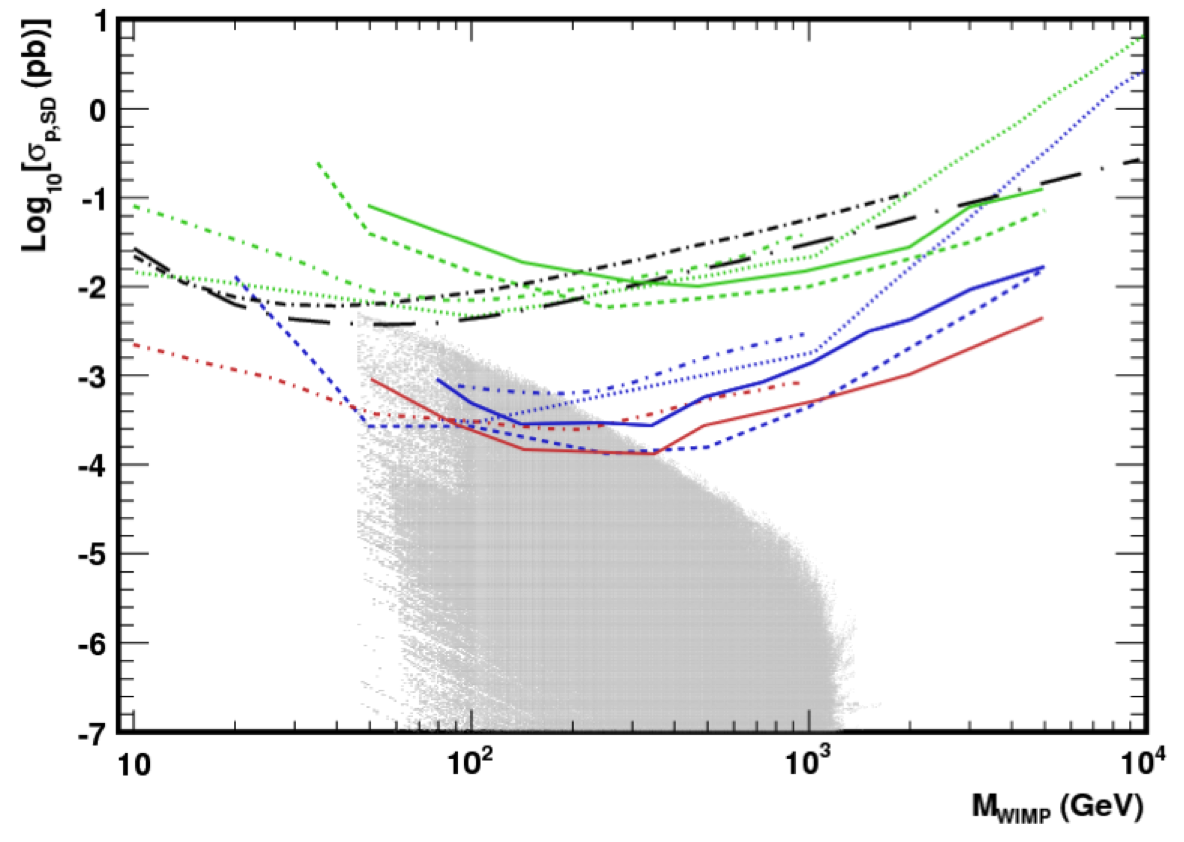} 
\includegraphics[width=0.5\linewidth]{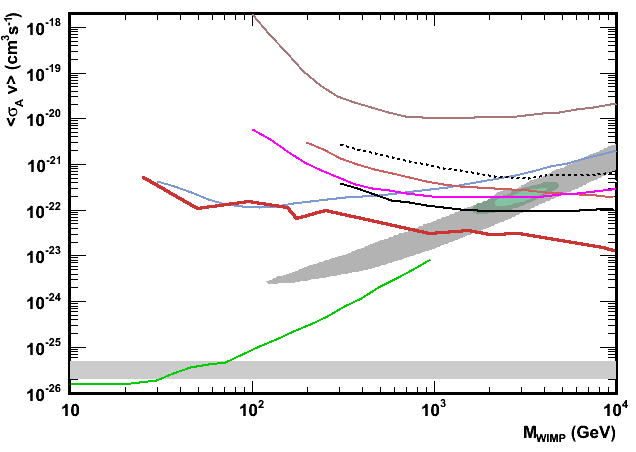}

\caption{Top: Spin-dependent cross-section limits for the search on the
  Sun: ANTARES 2007-2012 (thick solid lines):  $\tau^{+}\tau^{-}$
  (red), W$^{+}$W$^{-}$ (blue), b${\rm \bar{b}}$ (green), IceCube-79 (dashed lines), SuperKamiokande (colored
  dash-dotted lines), SIMPLE (black short dash-dotted line), COUPP
  (black long dash-dotted line) and XENON-100 (black long dashed
  line). The results are compared with a scan in
  MSSM-7. (Preliminary). Bottom: Limits on $<\sigma v >$ for
  the Galactic Centre for the $\tau^{+} \tau^{-}$ channel (red solid
  line) with IC40 for the GC (brown solid line), IC59 for dwarf
  galaxies (dashed black line), IC79 for the Galactic Halo (magenta solid
  line), IC59 for the VIRGO cluster (black solid line), DeepCore+IC79
  for the GC (blue solid line) and Fermi for dwarf galaxies (green
  solid line). The grey/green area represent leptophilic dark matter
  models which would explain the PAMELA (grey) and Fermi+PAMELA+HESS
  (green) excess in the
  Galactic Centre. (Preliminary).}
\label{sunlim}
\end{center}
\end{figure}

After unblinding the data, no excess of data over expected background
is found, so limits on the neutrino flux limit and on the WIMP-nucleon
cross section can be set (see~\cite{paper} for a previous search with
2007-2008 data). The results from the search can be seen in
Figure~\ref{sunlim} (top), where
the limits of the WIMP-proton scattering cross section are shown. It
is important to note that neutrino telescopes offer the best limits
for spin-dependent cross section (since the Sun is made basically of
protons), better than those from direct searches (other indirect
searches set limits on $<\sigma v>$).

\subsection{The Galactic Centre}
\label{sec:gc}

The Galactic Centre is also a very promising source for neutrino
telescopes. The distance is much larger, but the total mass involved
is also larger. Moreover, there is no absorption of neutrinos,
contrary to what happens in the Sun, which is particularly relevant
for high energy/masses, where the angular resolution improves. The signal is simulated using the simulation by Cirelli et
al.~\cite{cirelli}. The profile of the dark matter halo has been
simulated with the package CLUMPY. The background, as in
the case of the Sun, is evaluated by scrambling real data. After
unblinding the data, no significant excess is found so limits are set in
$<\sigma v>$, as shown in Figure~\ref{sunlim} (bottom) for a NFW halo profile
and the $\tau^{+} \tau^{-}$ channel. These limits exclude the
leptophilic dark matter interpretation of the 
Fermi+PAMELA+HESS excess. It can also be seen that above $\sim$150~GeV
the limits set by ANTARES are the best limits from neutrino
telescopes, since the visibility of ANTARES of the Galactic Centre and
its angular resolution are better than those of IceCube.


\subsection{The Earth}
\label{sec:earth}

\begin{figure}[c]
\begin{center}
\includegraphics[width=0.50\linewidth]{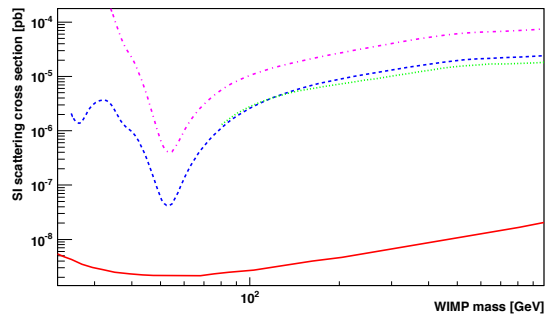}
\caption{Spin-independent cross section sensitivity (90\% CL) for the Earth
analysis, assuming $<\sigma v >$ $\sim$3x10$^{-26} cm^{-3}s^{-1}$ for three different
channels: $\tau^{+}\tau^{-}$ (dash, blue), W$^{+}$W$^{-}$ (dot, green) and
b${\rm \bar{b}}$ (dash-dot, magenta). This sensitivity is also compared with the
results of XENON-100 (solid, red). (Preliminary).}
\label{earthsen}
\end{center}
\end{figure}

Dark matter would also accumulate in the Earth after scattering, like
in the case of the Sun. However, in this case we cannot assume that an
equilibrium between caputre and annihilation has been
reached. Moreover, since the scattering is mostly on the heavy
elements in the Earth core, the limits are set on the spin-independent
cross section of WIMP scattering. These limits are particularly
interesting for WIMP masses close to the masses of scattering nuclei
(iron and nickel). The signal is evaluated with WIMPSim and the
background is calculated from the background in the zenith angle band
between 160$^{\circ}$ and 170$^{\circ}$ degrees ($>$95\% of the signal is found at zenith
$>$170$^{\circ}$). The sensitivity of ANTARES is shown in Figure~\ref{earthsen}.

\section{Conclusions}
\label{sec:conclusions}

The search for dark matter is one of the main scientific goals of
neutrino telescopes. ANTARES, installed in the Mediterranean Sea, has
been taking data since 2007. Several sources have been studied. In
this paper we have presented the results of the analysis for the Sun
and the Galactic Centre (in terms of limits) and the Earth (in terms
of sensitivity). Although no significant excess has been seen in the unblinded
data, competitive limits have been set.

\section*{Acknowledgments}

The authors acknowledge the financial support of the Spanish Ministerio de Ciencia e Innovaci\'on (MICINN), grants FPA2009-13983-C02-01, FPA2012-37528-C02-01, ACI2009-1020, Consolider MultiDark CSD2009-00064 and of the Generalitat Valenciana, Prometeo/2009/026.

\end{document}